\documentclass[%
  prl,showpacs,twocolumn,reprint,superscriptaddress,
  preprintnumbers,bibnotes,floatfix,numerical,graphics,nofootinbib]{revtex4-1}

\usepackage{graphics}
\usepackage{epsfig}

\usepackage{graphicx}
\usepackage{dcolumn}
\usepackage{bm}

\begin{document}

\title{Spin dependence of $\eta$ meson production 
in proton-proton collisions close to threshold}

\newcommand*{\IKPUU}{Division of Nuclear Physics, Department of Physics and 
 Astronomy, Uppsala University, Box 516, 75120 Uppsala, Sweden}
\newcommand*{\ASWarsN}{Department of Nuclear Physics, National Centre for 
 Nuclear Research, ul.\ Hoza~69, 00-681, Warsaw, Poland}
\newcommand*{\IPJ}{Institute of Physics, Jagiellonian University, prof.\ 
 Stanis{\l}awa {\L}ojasiewicza~11, 30-348 Krak\'{o}w, Poland}
\newcommand*{\Edinb}{School of Physics and Astronomy, University of Edinburgh, 
James Clerk Maxwell Building, Peter Guthrie Tait Road, Edinburgh EH9 3FD, 
Great Britain}
\newcommand*{\MS}{Institut f\"ur Kernphysik, Westf\"alische 
 Wilhelms--Universit\"at M\"unster, Wilhelm--Klemm--Str.~9, 48149 M\"unster, 
 Germany}
\newcommand*{\ASWarsH}{High Energy Physics Department, National Centre for 
 Nuclear Research, ul.\ Hoza~69, 00-681, Warsaw, Poland}
\newcommand*{\Budker}{Budker Institute of Nuclear Physics of SB RAS, 
 11~akademika Lavrentieva prospect, Novosibirsk, 630090, Russia}
\newcommand*{\Novosib}{Novosibirsk State University, 2~Pirogova Str., 
 Novosibirsk, 630090, Russia}
\newcommand*{\IKPJ}{Institut f\"ur Kernphysik, Forschungszentrum J\"ulich, 
 52425 J\"ulich, Germany}
\newcommand*{\IFJ}{The Henryk Niewodnicza{\'n}ski Institute of Nuclear 
 Physics, Polish Academy of Sciences, 152~Radzikowskiego St, 31-342 
 Krak\'{o}w, Poland}
\newcommand*{\PITue}{Physikalisches Institut, Eberhard--Karls--Universit\"at 
 T\"ubingen, Auf der Morgenstelle~14, 72076 T\"ubingen, Germany}
\newcommand*{\Kepler}{Kepler Center f\"ur Astro-- und Teilchenphysik, 
 Physikalisches Institut der Universit\"at T\"ubingen, Auf der 
 Morgenstelle~14, 72076 T\"ubingen, Germany}
\newcommand*{\ZELJ}{Zentralinstitut f\"ur Engineering, Elektronik und 
 Analytik, Forschungszentrum J\"ulich, 52425 J\"ulich, Germany}
\newcommand*{\Erl}{Physikalisches Institut, 
 Friedrich--Alexander--Universit\"at Erlangen--N\"urnberg, 
 Erwin--Rommel-Str.~1, 91058 Erlangen, Germany}
\newcommand*{\ITEP}{Institute for Theoretical and Experimental Physics named 
 by A.I.\ Alikhanov of National Research Centre ``Kurchatov Institute'', 
 25~Bolshaya Cheremushkinskaya, Moscow, 117218, Russia}
\newcommand*{\Giess}{II.\ Physikalisches Institut, 
 Justus--Liebig--Universit\"at Gie{\ss}en, Heinrich--Buff--Ring~16, 35392 
 Giessen, Germany}
\newcommand*{\IITI}{Department of Physics, Indian Institute of Technology 
 Indore, Khandwa Road, Simrol, Indore--453552, Madhya Pradesh, India}
\newcommand*{\HepGat}{High Energy Physics Division, Petersburg Nuclear Physics 
 Institute named by B.P.\ Konstantinov of National Research Centre ``Kurchatov 
 Institute'', 1~mkr.\ Orlova roshcha, Leningradskaya Oblast, Gatchina, 188300, 
 Russia}
\newcommand*{\HeJINR}{Veksler and Baldin Laboratory of High Energiy Physics, 
 Joint Institute for Nuclear Physics, 6~Joliot--Curie, Dubna, 141980, Russia}
\newcommand*{\Katow}{August Che{\l}kowski Institute of Physics, University of 
 Silesia, Uniwersytecka~4, 40-007, Katowice, Poland}
\newcommand*{\NITJ}{Department of Physics, Malaviya National Institute of 
 Technology Jaipur, JLN Marg Jaipur - 302017, Rajasthan, India}
\newcommand*{\JARA}{JARA--FAME, J\"ulich Aachen Research Alliance, 
 Forschungszentrum J\"ulich, 52425 J\"ulich, and RWTH Aachen, 52056 Aachen, 
 Germany}
\newcommand*{\Bochum}{Institut f\"ur Experimentalphysik I, Ruhr--Universit\"at 
 Bochum, Universit\"atsstr.~150, 44780 Bochum, Germany}
\newcommand*{\IITB}{Department of Physics, Indian Institute of Technology 
 Bombay, Powai, Mumbai--400076, Maharashtra, India}
\newcommand*{\Tomsk}{Department of Physics, Tomsk State University, 36~Lenina 
 Avenue, Tomsk, 634050, Russia}
\newcommand*{\KEK}{High Energy Accelerator Research Organisation KEK, Tsukuba, 
 Ibaraki 305--0801, Japan} 
\newcommand*{\ASLodz}{Department of Astrophysics, National Centre for Nuclear 
 Research, 90--950 {\L}\'{o}d\'{z}, Poland}

\author{P.~Adlarson}\altaffiliation[present address: ]{\Mainz}\affiliation{\IKPUU}
\author{W.~Augustyniak} \affiliation{\ASWarsN}
\author{W.~Bardan}      \affiliation{\IPJ}
\author{M.~Bashkanov}   \affiliation{\Edinb}
\author{S.~D.~Bass}     \affiliation{\IPJ}
\author{F.S.~Bergmann}  \affiliation{\MS}
\author{M.~Ber{\l}owski}\affiliation{\ASWarsH}
\author{A.~Bondar}      \affiliation{\Budker}\affiliation{\Novosib}
\author{M.~B\"uscher}\altaffiliation[present address: ]{\PGI, \DUS}\affiliation{\IKPJ}
\author{H.~Cal\'{e}n}   \affiliation{\IKPUU}
\author{I.~Ciepa{\l}}   \affiliation{\IFJ}
\author{H.~Clement}     \affiliation{\PITue}\affiliation{\Kepler}
\author{E.~Czerwi{\'n}ski}\affiliation{\IPJ}
\author{K.~Demmich}     \affiliation{\MS}
\author{R.~Engels}      \affiliation{\IKPJ}
\author{A.~Erven}       \affiliation{\ZELJ}
\author{W.~Erven}       \affiliation{\ZELJ}
\author{W.~Eyrich}      \affiliation{\Erl}
\author{P.~Fedorets}    \affiliation{\IKPJ}\affiliation{\ITEP}
\author{K.~F\"ohl}      \affiliation{\Giess}
\author{K.~Fransson}    \affiliation{\IKPUU}
\author{F.~Goldenbaum}  \affiliation{\IKPJ}
\author{A.~Goswami}     \affiliation{\IITI}\affiliation{\IKPJ}
\author{K.~Grigoryev}   \affiliation{\IKPJ}
\author{K.~Grigoryev}\affiliation{\IKPJ}\affiliation{\HepGat}
\author{C.--O.~Gullstr\"om}\affiliation{\IKPUU}
\author{L.~Heijkenskj\"old}\altaffiliation[present address: ]{\Mainz}\affiliation{\IKPUU}
\author{V.~Hejny}       \affiliation{\IKPJ}
\author{N.~H\"usken}    \affiliation{\MS}
\author{L.~Jarczyk}     \affiliation{\IPJ}
\author{T.~Johansson}   \affiliation{\IKPUU}
\author{B.~Kamys}       \affiliation{\IPJ}
\author{G.~Kemmerling}\altaffiliation[present address: ]{\JCNS}\affiliation{\ZELJ}
\author{G.~Khatri}\altaffiliation[present address: ]{\Harvard}\affiliation{\IPJ}
\author{A.~Khoukaz}     \affiliation{\MS}
\author{O.~Khreptak}    \affiliation{\IPJ}
\author{D.A.~Kirillov}  \affiliation{\HeJINR}
\author{S.~Kistryn}     \affiliation{\IPJ}
\author{H.~Kleines}\altaffiliation[present address: ]{\JCNS}\affiliation{\ZELJ}
\author{B.~K{\l}os}     \affiliation{\Katow}
\author{W.~Krzemie{\'n}}\affiliation{\IPJ}
\author{P.~Kulessa}     \affiliation{\IFJ}
\author{A.~Kup\'{s}\'{c}}\affiliation{\IKPUU}\affiliation{\ASWarsH}
\author{A.~Kuzmin}      \affiliation{\Budker}\affiliation{\Novosib}
\author{K.~Lalwani}     \affiliation{\NITJ}
\author{D.~Lersch}      \affiliation{\IKPJ}
\author{B.~Lorentz}     \affiliation{\IKPJ}
\author{A.~Magiera}     \affiliation{\IPJ}
\author{R.~Maier}       \affiliation{\IKPJ}\affiliation{\JARA}
\author{P.~Marciniewski}\affiliation{\IKPUU}
\author{B.~Maria{\'n}ski}\affiliation{\ASWarsN}
\author{H.--P.~Morsch}  \affiliation{\ASWarsN}
\author{P.~Moskal}      \affiliation{\IPJ}
\author{H.~Ohm}         \affiliation{\IKPJ}
\author{W.~Parol}       \affiliation{\IFJ}
\author{E.~Perez del Rio}\altaffiliation[present address: ]{\INFN}\affiliation{\PITue}\affiliation{\Kepler}
\author{N.M.~Piskunov}  \affiliation{\HeJINR}
\author{D.~Prasuhn}     \affiliation{\IKPJ}
\author{D.~Pszczel}     \affiliation{\IKPUU}\affiliation{\ASWarsH}
\author{K.~Pysz}        \affiliation{\IFJ}
\author{A.~Pyszniak}    \affiliation{\IKPUU}\affiliation{\IPJ}
\author{J.~Ritman}   \affiliation{\IKPJ}\affiliation{\JARA}\affiliation{\Bochum}
\author{A.~Roy}         \affiliation{\IITI}
\author{Z.~Rudy}        \affiliation{\IPJ}
\author{O.~Rundel}      \affiliation{\IPJ}
\author{S.~Sawant}      \affiliation{\IITB}
\author{S.~Schadmand}   \affiliation{\IKPJ}
\author{I.~Sch\"atti--Ozerianska}\affiliation{\IPJ}
\author{T.~Sefzick}     \affiliation{\IKPJ}
\author{V.~Serdyuk}     \affiliation{\IKPJ}
\author{B.~Shwartz}     \affiliation{\Budker}\affiliation{\Novosib}
\author{K.~Sitterberg}  \affiliation{\MS}
\author{T.~Skorodko}\affiliation{\PITue}\affiliation{\Kepler}\affiliation{\Tomsk}
\author{M.~Skurzok}     \affiliation{\IPJ}
\author{J.~Smyrski}     \affiliation{\IPJ}
\author{V.~Sopov}       \affiliation{\ITEP}
\author{R.~Stassen}     \affiliation{\IKPJ}
\author{J.~Stepaniak}   \affiliation{\ASWarsH}
\author{E.~Stephan}     \affiliation{\Katow}
\author{G.~Sterzenbach} \affiliation{\IKPJ}
\author{H.~Stockhorst}  \affiliation{\IKPJ}
\author{H.~Str\"oher}   \affiliation{\IKPJ}\affiliation{\JARA}
\author{A.~Szczurek}    \affiliation{\IFJ}
\author{A.~Trzci{\'n}ski}\affiliation{\ASWarsN}
\author{M.~Wolke}       \affiliation{\IKPUU}
\author{A.~Wro{\'n}ska} \affiliation{\IPJ}
\author{P.~W\"ustner}   \affiliation{\ZELJ}
\author{A.~Yamamoto}    \affiliation{\KEK}
\author{J.~Zabierowski} \affiliation{\ASLodz}
\author{M.J.~Zieli{\'n}ski}\affiliation{\IPJ}
\author{J.~Z{\l}oma{\'n}czuk}\affiliation{\IKPUU}
\author{P.~{\.Z}upra{\'n}ski}\affiliation{\ASWarsN}
\author{M.~{\.Z}urek}   \affiliation{\IKPJ}

\newcommand*{\Mainz}{Institut f\"ur Kernphysik, Johannes 
 Guten\-berg--Universit\"at Mainz, Johann--Joachim--Becher Weg~45, 55128 Mainz, 
 Germany}
\newcommand*{\PGI}{Peter Gr\"unberg Institut, PGI--6 Elektronische 
 Eigenschaften, Forschungszentrum J\"ulich, 52425 J\"ulich, Germany}
\newcommand*{\DUS}{\\and\\ Institut f\"ur Laser-- und Plasmaphysik, Heinrich--Heine 
 Universit\"at D\"usseldorf, Universit\"atsstr.~1, 40225 Düsseldorf, Germany}
\newcommand*{\JCNS}{J\"ulich Centre for Neutron Science JCNS, 
 Forschungszentrum J\"ulich, 52425 J\"ulich, Germany}
\newcommand*{\Harvard}{Department of Physics, Harvard University, 
 17~Oxford St., Cambridge, MA~02138, USA}
\newcommand*{\INFN}{INFN, Laboratori Nazionali di Frascati, Via E. Fermi, 40, 
 00044 Frascati (Roma), Italy}

\collaboration{WASA-at-COSY Collaboration}\noaffiliation

\date{\today}

\begin{abstract}
Taking advantage of the high acceptance and axial symmetry of the WASA-at-COSY detector,
and the high polarization degree of the proton beam of COSY, 
the reaction $\vec{p}p\to pp\eta$ has been measured close 
to threshold to explore the analyzing power $A_y$. 
The angular distribution of $A_y$ is determined 
with the precision improved by more than 
one order of magnitude with respect to previous results
allowing a first accurate comparison with theoretical
predictions.
The determined analyzing power is consistent with zero 
for an excess energy of $Q~=~15$~MeV signaling $s$ wave production 
with no evidence for higher partial waves.
At $Q~=~72$~MeV the data reveals strong interference of 
$Ps$ and $Pp$ partial waves and cancellation of $(Pp)^2$ 
and $Ss^{*}Sd$ contributions. 
These results rule out
the presently available theoretical predictions 
for the production mechanism of the~$\eta$ meson. 
\end{abstract}

\pacs{13.60.Le, 14.40.Aq}    
%
\maketitle

In recent decades hadron physics has been rich in discoveries 
in the low energy region where the interaction between 
hadrons is a manifestation of the strong force 
between their components
\cite{Krusche:2014ava,Moskal:2002jm,Bass:2015ova,Ericson:1988gk,
Hanhart:2003pg}. 
However there are still many open questions involving
the non-perturbative dynamics and details of hadron production processes.  
Spin observables offer an essential tool to yield new insight into this physics.
In this Letter we focus on $\eta$ meson production 
in low-energy proton-proton collisions with a 
polarized proton beam.
We report the first precise measurements of the 
analyzing power for the $\vec{p}p\to pp\eta$ reaction 
at two energies close to threshold.
These measurements yield new powerful 
 constraints on models of the $\eta$ production.

The presented results are based on about 200 times 
larger statistics
and drastically reduced systematic uncertainties
with respect to the previous experiments~\cite{Winter:2002ft,Czyzykiewicz:2006jb,Balestra:2004kg}.
The main improvement of systematics is due to:
     (i) axial symmetry and full acceptance of the 
WASA-at-COSY detector (more than 20 times larger than 
for the COSY-11 experiment~\cite{Brauksiepe:1996ii}),
    (ii) no magnetic field in the detector
and, in addition the systematics was controlled 
by the  measurements
    (iii) for  two spin orientations
and 
(iv) for two different decay channels of the $\eta$ meson.

Previous studies by the CELSIUS~\cite{Calen:1996mn,Calen:1997sf,Petren:2010zz,Calen:1998vh}, 
COSY~\cite{Smyrski:1999jc,AbdelBary:2002sx,Moskal:2003gt,Moskal:2009bd,Moskal:2008pi} 
and SATURNE~\cite{Chiavassa:1994ru,Hibou:1998de,Bergdolt:1993xc}
experiments of the total and differential cross-section for $\eta$ 
meson production 
in $pp$ and $pn$ collisions revealed that the $\eta$ 
meson is predominantly produced via the excitation of 
one of the nucleons to the $S_{11}$ current via exchange 
of virtual mesons with the subsequent decay into the 
proton-$\eta$ pair.  
This conclusion was obtained from the observation 
of a large $\eta$ production cross-section 
relative to the $\eta^{\prime}$ meson 
production and the isotropic angular distribution 
of the $\eta$ mesons  
in the center-of-mass system (CMS).    
Measurements of the total cross-section 
for $\eta$ production in different isospin channels 
\cite{Calen:1998vh,Moskal:2008pi}
revealed a strong contribution from isovector 
exchanges which is additive in proton-neutron 
collisions and which (partially) cancels in 
proton-proton collisions, 
bringing more constraints on theoretical models.
While much progress has been achieved,
the mechanism for the excitation of the colliding 
proton to a resonance state still remains very much 
incomplete with a host of models each with different
weighting of exchanges proposed to explain the dynamics.

Here we use spin as a tool to gain further insights.
The experiment involves a polarized proton beam with 
incident momentum in the $z$ direction and transversely polarized in the $y$ direction colliding with an 
unpolarized fixed proton target. 
The analyzing power is a sensitive extra constraint 
on the details of the $\eta$ production mechanism.

The $\eta$ meson production process proceeds through 
exchange of a complete set of virtual meson hadronic 
states, which in models is usually truncated 
to single-virtual-meson exchange.
Theoretical models have been proposed involving 
$\pi$, $\eta$, $\rho$, $\omega$ and $\sigma$ 
(correlated two-pion) exchanges 
\cite{Faldt:2001uz,Nakayama:2002mu,Nakayama:2003jn,Pena:2000gb,
Deloff:2003te,Shyam:2007iz}
and excited nucleon resonances,
primarily the $S_{11}(1535)$ 
plus small contributions from 
the $D_{13} (1520)$ and $P_{11} (1440)$ 
\cite{Nakayama:2002mu,Nakayama:2003jn}.
OZI-violating gluonic excitations might also couple to 
the flavour-singlet part of the $\eta$ meson in short 
distance proton-proton interaction \cite{Bass:1999is}.
These exchanges induce very different spin dependence in 
the production process.
Polarized beams and measurement of the analyzing power 
can therefore put powerful new constraints on theoretical
understanding of the $\eta$ production process.
For example, $\rho$ exchange and $\pi$ exchange models 
predict a near threshold analyzing power with different 
sign \cite{Faldt:2001uz,Nakayama:2003jn}. 
Isotropic (pure s-wave) production would give zero analyzing power.

The measurements of the $\vec{p}p\to pp\eta$ 
reaction were conducted by means of the large acceptance 
close to $4\pi$ 
and axially symmetric WASA-at-COSY spectrometer, operating as an internal fixed-target facility at the Cooler Synchrotron COSY~\cite{Maier:1997zj}. 
The vertically polarized proton beam was circulating through the vertical stream of the hydrogen pellets leading to the 
$\vec{p}p\to pp\eta$ reaction in the center of the WASA-at-COSY detector. 
The measurements were performed for the beam momentum values of 2026~MeV/c and 2188~MeV/c, 
corresponding  to excess energies in the CMS
of $Q~=~15$~MeV and 72~MeV, respectively. 
The orientation of the proton beam polarization was flipped for each accelerator cycle which lasted 90~s.

The charged hadronic ejectiles 
were registered by means of forward scintillator hodoscopes 
and the straw tube trackers and were identified using the energy deposited in the subsequent scintillator layers. 
The $\eta$ mesons were detected by the electromagnetic calorimeter and the plastic scintillator barrel. 
The production of the $\eta$ meson via 
the $\vec{p}p\to pp\eta$ reaction was identified 
using missing and invariant mass techniques. 
In total, more than 400000 $\eta$ meson events were 
identified and used for the determination of the analyzing power.

The center of the interaction region where the polarized proton beam collided with the pellet target 
was monitored with a precision of about 0.5~mm by the concurrent measurement of elastically scattered protons. 
The superconducting solenoid was switched off in order to minimize losses of the spin polarization. 
Only neutral decay products of the $\eta$ meson were reconstructed. 
In particular, the $\eta \to \gamma \gamma$ and 
$\eta \to 3\pi^0 \to 6\gamma$ decay channels with the highest branching ratios (altogether over 71\%~\cite{Olive:2016xmw}) 
were used in the presented analysis. 
A detailed description of the WASA-at-COSY experiment as well as methods and results of the monitoring of the interaction region
were given in the dedicated articles~\cite{Adam:2004ch,Bargholtz:2008aa,Adlarson:2015zta,Hodana:2014bua}.

The analyzing power $A_y(\theta_{\eta})$ for the given polar angle $\theta_{\eta}$ 
of the emission of the $\eta$ meson in the CMS
was determined from the asymmetry of the efficiency-corrected $\eta$ meson production yields $N_{\eta}$
\begin{equation}
Asymmetry(\theta_{\eta},\phi_{\eta}) \equiv  
\frac{ N_{\eta}(\theta_{\eta},\phi_{\eta}) - N_{\eta}(\theta_{\eta},\phi_{\eta}+\pi)}
     { N_{\eta}(\theta_{\eta},\phi_{\eta}) + N_{\eta}(\theta_{\eta},\phi_{\eta}+\pi)} 
\end{equation} 
extracted as a function of the azimuthal angle $\phi_{\eta}$
\begin{equation}
\label{eq:asymmetry}
Asymmetry(\theta_{\eta},\phi_{\eta})
 = P \cdot A_{y}(\theta_{\eta}) \cdot \cos(\phi_{\eta}),
\end{equation} 
where $P$ denotes the degree of the spin polarization of the proton beam.
In the presented analysis the Madison convention~\cite{Madison} was applied to fix the sign of the asymmetry.

\begin{figure}[t]
\mbox{\hspace{-0.3cm}\includegraphics[width=0.27\textwidth]{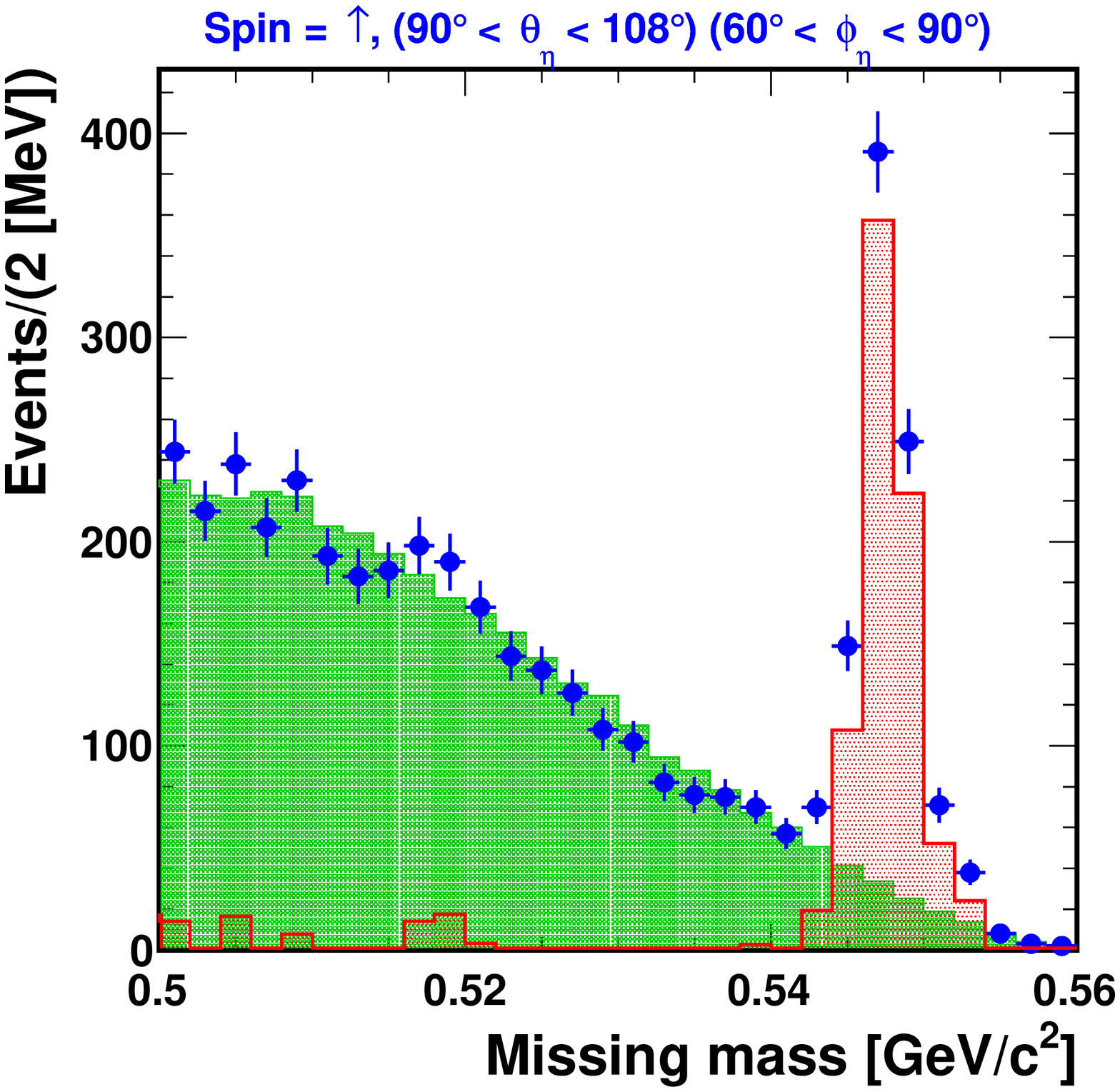}}%
\mbox{\hspace{-0.3cm}\includegraphics[width=0.27\textwidth]{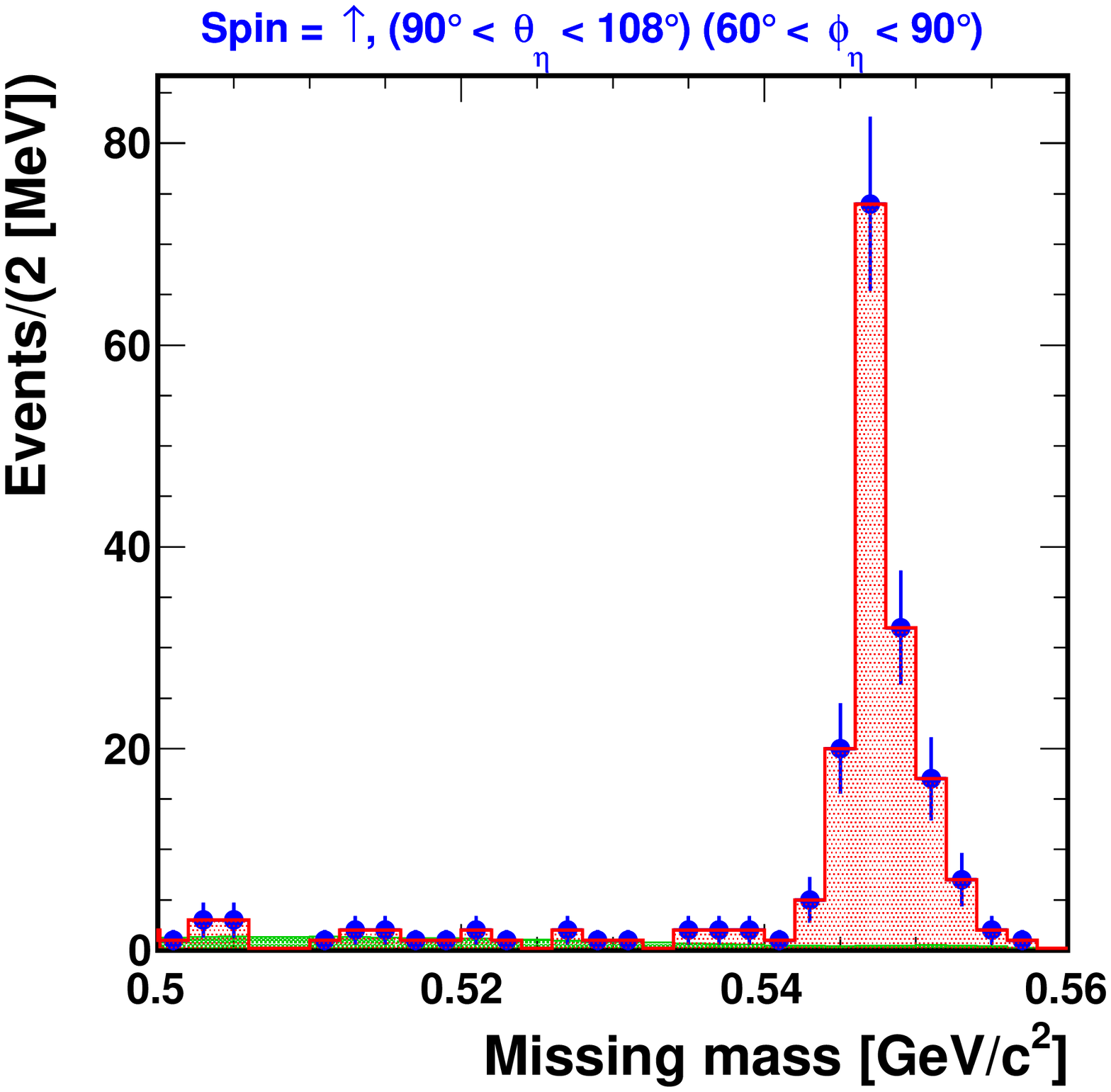}}%
\caption{\label{fig:mm} 
Examples of missing-mass distributions obtained for the 
excess energy Q~=~15~MeV:
(left) $\vec{p}p \to pp 2\gamma$, and 
(right) $\vec{p}p \to pp~3\pi^0 \to pp~6\gamma$ reactions.
The legends above the figures indicate the spin orientation and the angular intervals.
Experimental data are denoted by solid blue circles. 
Vertical bars indicate the statistical uncertainty. 
The shaded green area denotes the simulated contribution 
from multipion production background. 
The shaded red histograms corresponds to the $\eta$ events obtained by subtracting the multipion background.
}
\end{figure}

The yields of the $\eta$ meson production $N_{\eta}$ 
were determined based on the missing-mass spectra, independently for each ($\theta_{\eta}$,$\phi_{\eta}$) 
angular range.
Examples of spectra for a chosen angular range are 
presented in Fig.~\ref{fig:mm}. 
The spectra show only a small range of masses close to the kinematical limit 
where a clear signal from the production of the $\eta$ meson is seen on top of the multipion production background.
It is important to stress that 
the missing mass spectra for the $pp\to pp 2\gamma$ reaction channel shown in Fig.~\ref{fig:mm} (left)
include a background which is much larger when compared 
to the nearly negligible background observed for the $pp\to pp 6\gamma$ reaction shown in Fig.~\ref{fig:mm} (right)~\cite{Iryna}.
The contribution of the background was estimated by a 
fit to the experimental spectra of the shapes of missing 
mass distributions simulated for the production of 
the $\pi^0$, $2\pi^0$ and $3\pi^0$. 
The data simulated for the multipion production reaction included the response of the detector system 
and were analyzed using the same procedures as used for the analysis of the experimental events.
Next, after subtraction of the background the number 
of the registered $\eta$ mesons was
determined and corrected for the efficiency. 
The efficiency for the reconstruction of the $pp \to pp \eta$ reaction was established for each angular bin ($\theta_\eta, \phi_\eta$) separately based on the Monte-Carlo simulation performed taking into account the geometrical acceptance of the WASA-at-COSY detector as well as the experimental detection efficiencies and energy and angular resolutions.  The angular bin size (18 degrees for $\theta_\eta$ and 30 degrees for $\phi_\eta$) was chosen based on the statistical significance of events in each bin.

\begin{figure}[h]
\mbox{\hspace{-0.2cm}\includegraphics[width=0.50\textwidth]{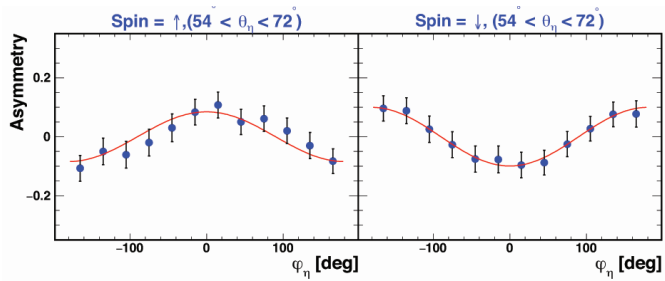}}%
\caption{\label{fig:Aypw} 
Example of asymmetry distributions as a function of 
the $\phi_{\eta}$ angle 
for a chosen $\theta_{\eta}$ angular bin, 
for an excess energy Q~=~72~MeV 
and the $\vec{p}p \to pp\eta \to 6\gamma$ reaction. 
Filled blue points denote extracted asymmetry values 
with the statistical uncertainty, while the red curves indicates the fit 
of Eq.~\ref{eq:asymmetry}  
to the experimental points.
The legend above the figures indicates spin orientation 
and the angular intervals.
The asymmetry for spin-up orientation is about 0.8 times smaller than the asymmetry for spin-down which is consistent with the ratio of the polarization values determined for different spin orientations (Table I).
}
\end{figure}

Figure~\ref{fig:Aypw} gives an illustration of our results for asymmetries for the $\vec{p}p \to pp \eta$ reaction yield as a function of the 
azimuthal angle $\phi_{\eta}$ of the $\eta$ meson momentum vector in the CM system.
A fit of Eq.~\ref{eq:asymmetry} 
to the angular dependence of the asymmetry
enables one to determine the product $P\cdot A_{y}$ 
which divided by the polarization $P$ gives the value 
of the analyzing power $A_y$.

The polarization $P$ was determined for each spin orientation and each excess energy separately,
by the simultaneous measurement of asymmetries for the elastically scattered protons 
for which the analyzing power is known~\cite{Bauer:1999vt,Altmeier:2000pe}. 
The method of the polarization analysis  
is described in detail in references~\cite{SchattiOzerianska:2015rwa,Ozerianska:2014qza,Iryna},
and the resulting values 
together with corresponding statistical uncertainties
are listed in Tab.~\ref{tab:table1}.
The systematic uncertainty of the polarization determination 
amounts to about 0.01 and, as it is shown in detail in 
Ref.~\cite{Hodana:2013cga},
it is predominantly due to the uncertainty of the reconstruction of the position of the interaction region. 
It is also important to stress that 
the spin polarization was stable during the whole run~\cite{SchattiOzerianska:2015rwa,Ozerianska:2014qza,Iryna}
and that, as expected, the analysis of the data taken with the unpolarized beam
resulted in the asymmetry equal to zero within the uncertainties~\cite{Ozerianska:2014qza,Iryna}.

\begin{table}[h]%
\caption{\label{tab:table1}%
The average polarization degree}
\begin{ruledtabular}
\begin{tabular}{lcc}
\textrm{$p_{beam}$~[MeV/c]}&
\textrm{Spin mode}&
\textrm{Polarization}\\
\colrule
2026 & up ($\uparrow$) & 0.793~$\pm$~0.010\\
     & down ($\downarrow$) & -0.577~$\pm$~0.007\\
2188 & up ($\uparrow$) & 0.537~$\pm$~0.009\\
     & down ($\downarrow$) & -0.635~$\pm$~0.011\\
\end{tabular}
\end{ruledtabular}
\end{table}

\begin{figure}[!h]
\mbox{\hspace{-0.6cm}\includegraphics[width=0.57\textwidth]{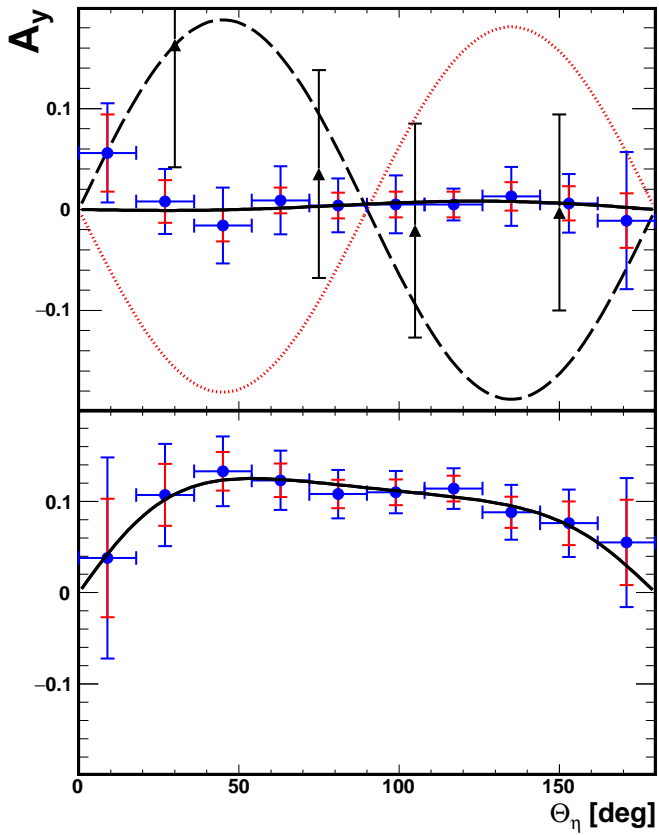}}
\caption{
{\it(upper panel) }
Analyzing power for the $\vec{p}p\to pp\eta$ reaction as a function of $\theta_{\eta}$ for Q~=~15~MeV.
The full circles represent the results obtained in this 
work, while the triangles are the values of the analyzing
powers measured by the COSY-11 Collaboration 
for Q~=10~MeV~\cite{Czyzykiewicz:2006jb}.
Horizontal error bars indicate the angular range. 
The vertical bars show total uncertainties with statistical 
and systematic errors separated by dashes.
The superimposed dotted line denotes the predictions based 
on pseudoscalar-meson-exchange model
\cite{Nakayama:2003jn}, 
whereas the dashed line represents 
the vector exchange model
 \cite{Faldt:2001uz}. 
The result of the fit of Eq.(4) to the data 
is presented by the solid line.
{\it(bottom panel)}
Analyzing power for the $\vec{p}p\to pp\eta$ reaction for the Q~=~72~MeV. 
}
\label{fig:AyQ15}

\end{figure}

Calculations of the analyzing power $A_y$ were conducted separately 
for spin-up and spin-down orientation,
and for each spin mode the $A_y$ was determined separately for two decay channels: $\eta \to 2\gamma$ and $\eta \to 6\gamma$. 
Moreover, for all above cases, the number of events $N_\eta$ corresponding to the $\vec{p}p \to pp\eta$ reactions, 
has been determined for each angular bin $(\theta_{\eta},\phi_{\eta})$ separately.
This enabled us to control systematic uncertainties 
which may occur due to the misalignment of the detector
and due to the methods of event reconstruction. 
The final results obtained by averaging values determined 
for both spin orientations and both decay channels
are given in Tab.~\ref{tab:TabAy} and are presented in 
Fig.~\ref{fig:AyQ15}.
Due to the axial symmetry of the detector, any unknown detector asymmetries and unknown efficiencies should cancel when averaging results obtained for two opposite spin orientations. But, anyhow the differences in these results were taken into account in the estimation of the systematic errors.
The systematic uncertainties listed in the table have been estimated by calculating changes in the values of $A_y$ 
to the variation of the parameters used in the analysis. After changing a tested parameter the full analysis chain was repeated and new $A_y$ values were determined.
In particular, the following contributions to the systematic error were taken into account~\cite{Iryna}: 
(i) selection criteria used in the particle identification, 
(ii)  range in the missing mass spectra used for counting the number of produced $\eta$ mesons, 
(iii) differences between $A_y$ values obtained for different decay channels
(iv) uncertainty of the values of polarization, and 
(v) differences between $A_y$ values obtained for spin-up and spin-down measurements.
The largest impact on the systematic error comes from the $A_y$ measurement combining different decay channels of the $\eta$ meson.

Integrating over the proton degrees of freedom
results in the analyzing power $A_y(\theta_{\eta})$, 
which in a partial-wave decomposition may be expressed as follows:
\begin{eqnarray}
& & A_y(\theta_{\eta}) \frac{d \sigma}{d \Omega_{\eta}}  
= 2 \pi [ G_1^{y0} \sin \theta_{\eta}
+ (H_1^{y0} + I^{y0}) \sin 2 \theta_{\eta}].
\end{eqnarray}
Here the form factors 
$G_1^{y0}$, $H_1^{y0}$ and $I^{y0}$ are defined
in Winter et al. 
\cite{Winter:2002ft},
which generalizes
the analysis of spin dependence of $\pi^0$ 
production with polarized proton beams to $\eta$ production 
\cite{Meyer:2001gj}.
The superscript $y0$ indicates beam 
polarization along the $y$ axis and an unpolarized target.

We apply the usual spectroscopic notation to
describe the $pp \rightarrow pp \eta$ process,
viz.
$^{2S^i+1}L^i_{J^i} \rightarrow ^{2S+1}L_J, l 
$. 
Here, the relative orbital angular momentum of 
the two outgoing protons in their rest frame is 
denoted by capital
letters $L_p$ = S, P, D, ..., the one of the $\eta$ meson in
the CMS by the small letters $l_q$ = s,p,d, ...
With this notation
the individual terms
$G_1^{y0}$, $H_1^{y0}$ and $I^{y0}$ correspond to 
(Ps$^{*}$Pp), (Pp)$^2$ and (Ss$^{*}$Sd) interference, respectively.
The Pauli principle means that even and odd partial 
waves of the protons in the final state cannot interfere 
with each other.
In total, with polarized beams there are 
one Ss, two Ps, nine Pp and two Sd final-state
production amplitudes \cite{Meyer:2001gj}.
For example, the Ss amplitude corresponds
to the process $^3\!P_0 \rightarrow ^1\!\!S_0, s$.
The (Pp)$^2$ and Ss$^{*}$Sd interference amplitudes 
always appear together in the analyzing power
and angular distribution~\cite{Meyer:2001gj}.


\begin{table}[h]%
\caption{\label{tab:TabAy}%
Analyzing power $A_y$ with statistical and systematic uncertainties
detemined for the $\vec{p}p\to pp\eta$ reaction at
excess energies of Q~=~15~MeV and Q~=~72~MeV.}
\begin{ruledtabular}
\begin{tabular}{ccc}
\textrm{$\theta_{\eta}$~[deg]}&
\textrm{$A_y(\theta_{\eta})$ for Q~=~15~MeV}&
\textrm{$A_y(\theta_{\eta})$ for Q~=~72~MeV}\\
\colrule
0~-~18  & 0.056~$\pm$~0.038~$\pm$~0.011  & 0.038~$\pm$~0.065~$\pm$~0.045\\
18~-~36  & 0.008~$\pm$~0.021~$\pm$~0.011 & 0.107~$\pm$~0.034~$\pm$~0.022\\
36~-~54  &-0.016~$\pm$~0.016~$\pm$~0.022 & 0.133~$\pm$~0.021~$\pm$~0.017\\
54~-~72  & 0.009~$\pm$~0.013~$\pm$~0.021 & 0.123~$\pm$~0.018~$\pm$~0.014\\
72~-~90  & 0.004~$\pm$~0.013~$\pm$~0.014 & 0.108~$\pm$~0.016~$\pm$~0.011\\
90~-~108 & 0.005~$\pm$~0.013~$\pm$~0.046 & 0.110~$\pm$~0.014~$\pm$~0.009\\
108~-~126 & 0.005~$\pm$~0.013~$\pm$~0.003& 0.114~$\pm$~0.014~$\pm$~0.008\\
126~-~144 & 0.013~$\pm$~0.014~$\pm$~0.015& 0.088~$\pm$~0.017~$\pm$~0.013\\
144~-~162 & 0.006~$\pm$~0.017~$\pm$~0.012& 0.076~$\pm$~0.024~$\pm$~0.013\\
162~-~180 &-0.011~$\pm$~0.027~$\pm$~0.041& 0.055~$\pm$~0.047~$\pm$~0.024\\\colrule
\end{tabular}
\end{ruledtabular}
\end{table}

Following Eq.(3) we fit the data as
\begin{equation}
A_y(\theta_{\eta}) \frac{d \sigma}{d \Omega_{\eta}} 
 = 
C_1 \sin \theta_{\eta} 
+ 
C_2 
\cos \theta_{\eta} \sin \theta_{\eta}
,
\end{equation}
where $C_1$ and $C_2$ are treated as free parameters.
For $Q~=~15$~MeV the angular distribution of the cross-section $\frac{d \sigma}{d \Omega_{\eta}}$ 
was assumed to be constant as determined by the COSY-11~\cite{Moskal:2003gt} 
and COSY-TOF~\cite{AbdelBary:2002sx} experiments.
For the fit at $Q~=~72$~MeV the 
$\frac{d \sigma}{d \Omega_{\eta}}$ determined by 
the WASA-CELSIUS collaboration~\cite{Petren:2010zz} was used.
For $Q=15$ MeV we find 
$C_1 =  (0.001 \pm 0.001)~\mu b/sr$ 
and
$C_2 = (-0.002 \pm 0.003)~\mu b/sr$. 
For $Q=72$ MeV we obtain the fit parameters:
$C_1 = (0.104 \pm 0.006)~\mu b/sr$ 
and
$C_2 = (0.020 \pm  0.012)~\mu b/sr$.

At $Q=15$ MeV we find no evidence for partial waves beyond
s-wave production.
At $Q=72$ MeV we find evidence for a significant contributions of 
 higher partial waves.
If we conclude from the finite coefficient $C_1$ of 
the $\sin \theta_{\eta}$ term that both Ps and Pp 
give significant contributions,
then the vanishing (within errors) coefficient $C_2$ 
points to a cancellation between (Pp)$^2$ and Ss*Sd
contributions.

Previously, in the 
COSY-11 analysis of the 15.5 MeV M(pp) shape it was 
suggested that the high-mass region was a signal for 
a Ps contribution at $Q=15$ MeV \cite{Moskal:2003gt}. 
If this is indeed present in the data, 
then the small coefficient $C_1$ would indicate a small
Pp contribution at this excess energy.
At $Q=72$ MeV Petren et al \cite{Petren:2010zz}
found that a sizable Pp contribution is needed to get 
the valley along the diagonal of the Dalitz plot
for the $pp \rightarrow pp \eta$ reaction. 
Maximal Ss*Sd interference there was suggested to 
explain the angular distribution of $\eta$ production
at $Q=40$ MeV.

All together, these previous results and the ones presented 
in this Letter have a following interpretation.

First, the data indicate just s-wave production at $Q=15$ MeV.
This result contradicts predictions based on single meson exchange as shown in Fig.~\ref{fig:AyQ15}.

Measurements of the isospin dependence of $\eta$ meson production in proton-nucleon collisions 
have revealed that the total cross-section 
for the quasi-free $pn \rightarrow pn \eta$ exceeds a corresponding cross-section for $pp \rightarrow pp \eta$ 
by a factor of about three at threshold and by a factor of 
six at higher excess energies between about 25 and 100 MeV~\cite{Calen:1998vh,Moskal:2008pi}. 
This isospin dependence is interpreted as evidence 
for a strong isovector exchange contribution which 
exhibits (partial) cancellation in proton-proton
collisions and addition in proton-neutron collisions.
This isovector exchange was interpreted in terms of 
the $\rho$ meson in Ref.~\cite{Faldt:2001uz} and $\pi$ 
exchange in~\cite{Nakayama:2003jn}.
These one-boson exchange models, when fit to early data 
on $\eta$ production, made predictions for $A_y$ as 
shown in Fig.~\ref{fig:AyQ15} with 
$A_y (\theta_{\eta}) = {\cal A} \sin 2 \theta_{\eta}$
where $|{\cal A}| =0.18$ at $Q=15$ MeV.
Note here that these $\rho$ \cite{Faldt:2001uz} and 
$\pi$ \cite{Nakayama:2003jn} exchange curves come 
with the opposite sign, i.e.\ the distribution is 
shifted by $\theta_{\eta} = 90$ degrees.
One possible explanation might be cancellation through
destructive interference between $\pi$ and $\rho$ 
exchanges in $\eta$ production in proton-proton 
collisions very close to threshold together with a strong (spin-independent) scalar $\sigma$ exchange contribution.

Cancellation of (Pp)$^2$ and Ss*Sd interference terms at 
$Q=72$ MeV suggests a phase cancellation of various 
meson exchanges and resonance contributions, 
e.g. associated 
with the nucleon resonances
$S_{11}(1535)$, $D_{13} (1520)$ and $P_{11} (1440)$.

To summarize, we have measured the spin analyzing power 
for $\eta$ production close to threshold with 
precision improved by one order of magnitude. 
For excess energy $Q=15$ MeV the data is consistent 
with the $\eta$ production process in pure s-wave.
We find evidence of higher partial waves at $Q=72$ MeV.
The Ps$^{*}$Pp interference determines the shape of the measured analyzing power with cancellation of Ss*Sd and 
(Pp)$^2$ interference terms.
The data contradicts predictions of presently available
meson exchange models of the production mechanism.
The analyzing power complements previous measurements
of the energy and angular distribution of $\eta$ meson
production and 
provides new valuable constraints for future model building.

\begin{acknowledgments}
We thank C. Wilkin for helpful communications.
This work was supported 
by the Polish National Science Centre through the Grants
No. 
2013/11/N/ST2/04152,   
2014/15/N/ST2/03179
and 2016/23/B/ST2/00784, 
and the Forschungszentrum
J\"{u}lich FFE Funding Program of the J\"{u}lich Center 
for Hadron Physics, and the Swedish Research Council.

\end{acknowledgments}

\bibliography{PRL_Ayw-referee}
%
\end{document}